\documentstyle[aps,eqsecnum,epsf]{revtex}

\begin{document}
\thispagestyle{empty}

\vspace{7mm}
\begin{center}
{\Large{\bf Dispersion relations and the $\Delta$ contributions
into the amplitudes $M_{1+}^{3/2},E_{1+}^{3/2}$ 
from new VPI partial-wave analysis of pion photoproduction}}\\
\vspace{7mm}
{\large I.G.Aznauryan}\\
\vspace{7mm}
{\em Yerevan Physics Institute,
Alikhanian Brothers St.2, Yerevan, 375036 Armenia}\\
{(e-mail addresses: aznaur@jerewan1.yerphi.am, aznaury@cebaf.gov)}\\
\vspace{7mm}

\end{center}

\begin{abstract}
Within fixed-t dispersion relations
the results of new VPI partial-wave analysis for
the multipole amplitudes
$M_{1+}^{3/2},E_{1+}^{3/2}$
are successfuly described,
and the resonance and nonresonance contributions
into these amplitudes are separated
in correspondence with the interpretation
on the language
of diagram approach, dynamical models
and effective Lagrangian approach.
The amplitudes $A_p^{3/2}$ and $A_p^{1/2}$
corresponding to the $\Delta$
contributions into $M_{1+}^{3/2},E_{1+}^{3/2}$
are obtained. They are in better agreement with 
quark model predictions than the amplitudes
extracted without subtraction of the nonresonance contributions
in $M_{1+}^{3/2},E_{1+}^{3/2}$. 
The obtained value of the ratio $E2/M1$
for the $\gamma N \rightarrow P_{33}(1232)$
transition is: $E2/M1=-0.022\pm 0.004$.
\end{abstract}

\vspace{3mm}

\section{Introduction}

It is known that the investigation of the transition
$\gamma N \rightarrow P_{33}(1232)$,
using the experimental data on the pion
photoproduction on the nucleons,
is connected with the problem of separation
of the resonance and nonresonance contributions
in the multipole amplitudes $M_{1+}^{3/2},E_{1+}^{3/2}$,
which carry information on this transition.
These amplitudes may contain significant nonresonance
contributions, the fact which was clear
with obtaining the first accurate data \cite{1,2} on the 
amplitude $E_{1+}^{3/2}$.
The energetic behaviour of this amplitude,
in fact, is incompatible with
the resonance behaviour.
The first investigations of this problem \cite{3,4,5} have shown
that it is closely related to the problem
of fulfilment of the unitarity condition,
which for photoproduction multipole amplitudes
(let us denote them as $M(W)$)
in the $P_{33}(1232)$
resonance region means the fulfilment
of the Watson theorem\cite{6}:
\begin{equation}
M(W)=
\exp[i\delta(W)]
|M(W)|.
\label{1}\end{equation}
Here $\delta$ is the phase of the
corresponding $\pi N$ scattering amplitude:
\begin{equation}
h(W)=\sin[\delta(W)]\exp[i\delta(W)].
\label{2}\end{equation}

There are different approaches for the extraction 
of an information on the $\gamma N \rightarrow P_{33}(1232)$
transition from the pion photoproduction data 
with different forms of the unitarization of the
multipole amplitudes.
These approaches  can be subdivided into the following
groups: the phenomenological approaches \cite{3,4,5,7}
including the approaches based on the K-matrix
formalism \cite{8,9}, the effective Lagrangian (EL) approaches
\cite{10,11,12,13,14} with different phenomenological
form of the unitarization of the amplitudes, 
the dynamical models (DM) \cite{15,16,17,18,19,20,21},
and the approaches based on the fixed-t
dispersion relations \cite{22,23,24,25}.

In Refs. \cite{24,25} it was shown
that fixed-t  dispersion relations
used within the approach of Refs. \cite{26,27}
can be usefull for the separation of the resonance
and nonresonance contributions in the 
amplitudes $M_{1+}^{3/2},E_{1+}^{3/2}$.
In Sec.2 we specify this separation,
making correspondence between the contributions
in EL approaches and DM
and the solutions of the integral equations
for  $M_{1+}^{3/2},E_{1+}^{3/2}$,
which follow from dispersion relations
for these amplitudes within the approach
of Refs. \cite{26,27}. These solutions are used
in Sec.3 as the input for the description
of the results of VPI partial-wave analysis \cite{28}
for $M_{1+}^{3/2},E_{1+}^{3/2}$,
and for separation of the resonance and nonresonance 
contributions in these amplitudes.
\section{Correspondence between  contributions
in dispersion relation, effective Lagrangian 
and dynamical approaches}

The results of EL approaches and DM
can be interpretated on the diagram language, which
is most suitable for comparison with the predictions
of existing models, because current hadron models
and approaches (quark model, bag model, QCD sum rules ...)
operate only with verteces and can not predict
the whole amplitudes of the processes. 
In EL approaches and DM the amplitudes
$M_{1+}^{3/2},E_{1+}^{3/2}$ are described
in terms of the diagrams corresponding to the
$N$ exchange in the u-channel, $\Delta$ exchange in the $s$- and $u$-
channels, and $\pi$ and $\omega$ exchanges
in the $t$- channel. The proper phase of the amplitudes 
is obtained via taking into account final state interaction.
This procedure is carried out in EL models phenomenologicaly
using the Olsson \cite{3}, Noelle \cite{29}
and K-matrix approaches. In DM the unitarization
of the amplitudes is made within some method for calculation
of the diagrams corresponding to the final state interaction.
These calculations are made using different approaches
for fulfilment of relativistic and gauge invariance
with different methods of cutoff and incorporation
of the off-shell effects in the integrals.

The amplitudes corresponding to the $N,~\pi$ and $\omega$
exchanges and to the $\Delta$ exchange in the $u$-
channel are real. Let us denote their contribution
into $M_{1+}^{3/2}$ and $E_{1+}^{3/2}$
as $M^{NR}(W)$. In the quantum mechanics,
rescattering effects in these amplitudes
lead to the following replacement
(see Ref.\cite{30}, Chapter 9):
\begin{equation} 
M^{NR}\rightarrow M^{NR}_{rescat}=M^{NR}+
\frac{1}{\pi}\frac{1}{D(W)}
\int\limits_{W_{thr}}^{\infty}
\frac{D(W')h(W')M^{NR}(W')}{W'-W-i\varepsilon}dW'=
\label{3}\end{equation} 
\begin{equation}
=\exp [i \delta (W)]\left[M^{NR}(W)\cos \delta (W)+
e^{a(W)}r(W)\right],
\label{4}\end{equation}
where
\begin{equation}
r(W)=
\frac{P}{\pi}
\int\limits_{W_{thr}}^{\infty}
\frac{e^{-a(W')}\sin \delta (W')M^{NR}(W')}{W'-W}dW',
\label{5}\end{equation}
\begin{equation}
a(W)=
\frac{P}{\pi}
\int\limits_{W_{thr}}^{\infty}
\frac{W \delta (W')}{W'(W'-W)}dW'.
\label{6}\end{equation}

In Eq.(\ref{3}) it is supposed that the unitarity
condition  (\ref{1}) can be used
in the whole range of integration,
$D(W)$ is the Jost function:
\begin{equation}
1/D(W)=
\exp\left[\frac{W}{\pi}
\int\limits_{W_{thr}}^{\infty}
\frac{\delta (W')}{W'(W'-W-i\varepsilon)}dW'\right]=
\exp [i \delta (W)]e^{a(W)}.
\label{7}\end{equation} 

The contributions analogous to both terms
in Eqs. (\ref{3}),(\ref{4}) exist in all DM. These models
reproduce exactly the first term in Eq. (\ref{4}),
second one being 
model dependent and different in different models.
In EL approaches the unitarization made via the Noelle
and K-matrix ansatzes corresponds to taking
into account only the first term in  Eq. (\ref{4})
(see Ref. \cite{13}). The unitarization
via the Olsson ansatz in these approaches
has no analogy with the above formulas.
It is interesting that just the first term
in Eq. (\ref{4}) determines the nonresonance
behaviour of the multipole amplitude $E_{1+}^{3/2}$
(see below the curve 5 in Fig. 2). 

In the absence of background contribution
into $\delta_{1+}^{3/2}$,
incorporation of the $\pi N$ rescattering
in the resonance parts ($M^R$) of the amplitudes
$M_{1+}^{3/2},E_{1+}^{3/2}$
leads in the vicinity of $\Delta$ to the following replacements:
\begin{equation} 
M^{R}=
\frac{f_{\pi N,\Delta}^0 f_{\Delta,\gamma N}^0} 
{s -m_{0\Delta}^2}
\rightarrow 
\frac{f_{\pi N,\Delta} f_{\Delta,\gamma N} }
{s -m_{\Delta}^2-im_{\Delta}\Gamma_{\Delta}}
\equiv \frac{f_{\pi N,\Delta} f_{\Delta,\gamma N}}
{m_{\Delta}\Gamma_{\Delta}}
sin\delta_R e^{i\delta_R}.
\label{8}\end{equation}

Here $f_{\pi N,\Delta},~ f_{\Delta,\gamma N},~\Gamma_{\Delta}$
and $m_{\Delta}$ are dressed verteces and
$\Delta$ width and mass; the corresponding values
containing "0" are bare ones.

The modification of $M^R$, due to the presence 
of the background contribution
in $\delta_{1+}^{3/2}$, can be taken
into account only phenomenologicaly.
One can estimate the magnitude of this modification
using the results of Ref.\cite{13} 
obtained within the Noelle and K-matrix forms
of the unitarization of the  amplitudes.
At the resonance position, where $\delta_{1+}^{3/2}=90^{\circ}$
(for the phase shift analysis of Refs. \cite{31,32} it is $W_R=1.229~GeV$),
the unitarization of $M^R$ within these methods
leads to the same results; namely, in  Eq. (\ref{8})
the replacement $e^{i\delta_R}\rightarrow e^{i\delta_{1+}^{3/2}}$
should be made, $sin\delta_R$ being equal to 1
in the K- matrix approach and to 0.97 in the Noelle approach.
This difference in $3\% $ we will consider as the uncertainty
of $M^R$ coming from the incorporation
of the background contribution into 
$\delta_{1+}^{3/2}$ at $W=W_R$.

Let us turn now to the dispersion relations.
Dispersion relations for  multipole amplitudes
follow from dispersion relations
for invariant amplitudes, defined in accordance
with  the hadron
current, which obeys the requirements
of the relativistic and gauge invariance
and the crossing invariance under 
the replacement $s \leftrightarrow u$.
Let us write these dispersion relations
for the multipole amplitudes
$M_{1+}^{3/2},E_{1+}^{3/2}$ in the form:
\begin{equation} 
M(W)=M^B(W)+M^{high}(W)+
\frac{1}{\pi}\int\limits_{W_{thr}}^{W_{max}}
\frac{h^*(W') M(W')}{W'-W-i\varepsilon}dW'+
\frac{1}{\pi}\int\limits_{W_{thr}}^{W_{max}}
K(W,W')h^*(W') M(W')dW',
\label{9}\end{equation}
where we have divided dispersion integrals
into two parts: from threshold up to $W_{max}=1.55~GeV$
(the region which is dominated by the $\Delta$ contribution),
and from $W_{max}$ up to $\infty$. 
Such division of the dispersion integrals,
with the consideration of $M^{high}(W)$
as a nonsingular function,
is possible only in the case,
if $h(W)\rightarrow 0$ when $W\rightarrow W_{max}$.
This condition was not taken into
account in Ref. \cite{23}. By this reason
the solutions of integral equations,
obtained in \cite{23},  are divergent at $W\rightarrow W_{max}$.

For the multipole amplitudes
$M_{1+}^{3/2},E_{1+}^{3/2}$ 
one can introduce the condition:
$h(W)\rightarrow 0$ when $W\rightarrow W_{max}$,
at $W_{max}\simeq1.55~GeV$,
because at $W=1.5~GeV$ we have $\delta_{1+}^{3/2}=164^{\circ}$\cite{31,32}.
From the $\pi N$ phase shift analyses (see, for example, \cite{31,32})
it is known that the amplitude $h_{1+}^{3/2}$
is elastic in the first integration region;
by this reason in the integrals
over the region $(W_{thr},W_{max})$
the imaginary parts of the multipole amplitudes
are written in the form:
$Im~M(W)=h^*(W)M(W)$, which follow from Eqs.(\ref{1},\ref{2}).
Therefore, the dispersion relations (\ref{9})
for the amplitudes
$M_{1+}^{3/2},E_{1+}^{3/2}$
can be considered as integral equations for these
amplitudes in the region $(W_{thr},W_{max})$.

In Eq. (\ref{9}), $M^B(W)$  is the Born term which
corresponds to the $N$ and $\pi$ exchanges,
with pseudoscalar coupling for the $NN\pi$ vertex.
The corresponding term in EL approaches
and DM is obtained  using pseudovector
coupling for this vertex; it differs from the
Born contribution by the nonsingular term which 
contributes only into the $B_1^{(+,0)}$ Ball amplitude:
\begin{equation} 
B_1^{(+,0}(s,t)=\frac{ge}{4m_N^2}g^{(v,s)},
\label{10}\end{equation}
where $m_N$ is the nucleon mass, and
\begin{equation} 
e^2/4\pi=1/137,~g^2/4\pi=14.5,~g^{(v)}=3.7,~g^{(s)}=-0.12.
\label{11}\end{equation}

The contribution of this term into our final results
is negligibly small.
$K(W,W')$  is a nonsingular
kernel arising from  the $u$- channel contribution
into the dispersion integral and the nonsingular
part of the $s$- channel contribution.
In the integrand of the relation (\ref{9}),
we did not write the couplings of
$M(W)$ to other multipoles;  by our estimations
their contributions into our final results are negligibly
small.

The values of the high energy integrals in Eq. (\ref{9})
can be evaluated using the results of
analyses of pion photoproduction on nucleons at high energies.
In our estimations we have used the results obtained in Ref. \cite{33},
where different variants of the description
of these data are considered within the approach based on the 
Regge poles and cuts. Our estimations have shown
that the high energy integrals
in Eq. (\ref{9}) can be roughly
approximated by the $\omega$ exchange, which contributes to the following 
Ball amplitudes:
\begin{equation} 
B_6^{(+)}=\frac{2g_{\gamma\omega\pi}g_{\omega NN}}{t-m_{\omega}^2},
~B_1^{(+)}=m_NB_6^{(+)},
\label{12}\end{equation}
where $m_{\omega}$ is the $\omega$ mass,
and $g_{\gamma\omega\pi}$ is related to
the $\omega\rightarrow\pi\gamma$ decay width by:
\begin{equation}
\Gamma(\omega\rightarrow\pi\gamma)=\frac
{g_{\gamma\omega\pi}^2k^3}{12\pi},
\label{13}\end{equation}
$k$ is the pion 3-momentum in the $\omega$
rest frame. From the data on $\Gamma(\omega\rightarrow\pi\gamma)$
\cite{34} we get $g_{\gamma\omega\pi}=0.73~GeV^{-1}$.
In Eq. (\ref{12}) we have presented only the contribution 
corresponding to the vector coupling in the vertex $\omega NN$,
because the role of the tensor $\omega NN$ coupling in our final
results is negligibly small.
For the vector coupling constant we have:
$g_{\omega NN}=8-14$  \cite{35}.
The results presented below in Figs. 1,2
correspond to the mean value of $g_{\omega NN}$
in this interval.

At $K(W,W')=0$, the integral equation (\ref{9})
has a solution in an analitical form
(see Refs.\cite{26,27} and the refferences therein):
\begin{equation}
M_{K=0}(W)=M_{part,K=0}^{B,\omega}(W)+c_MM_{K=0}^{hom}(W).
\label{14}\end{equation}

Here $M_{part,K=0}^{B,\omega}(W)$ is the
particular solution of Eq. (\ref{9})
generated by $M^B$ and $M^{\omega}$.
It is described by Eq. (\ref{3})
with the replacement $M^{NR}\rightarrow M^B+M^{\omega}$.
With this, in all integrals of Eqs.(\ref{3})-(\ref{7})
at $W'>W_{max}$, one should take $\delta(W')=\pi$.
So, $M_{part,K=0}^{B,\omega}(W)$
reproduces the nonresonance contributions
into the amplitudes $M_{1+}^{3/2},E_{1+}^{3/2}$,
generated by the $N,~\pi$ and $\omega$ exchanges,
when the final state interaction, caused by the $\pi N$
rescattering in the $\Delta$ region,
is taken into account in accordance with Eq. (\ref{3}).

$M_{K=0}^{hom}(W)=1/D(W)$ is the solution of the homogeneous
equation, which follow from (\ref{9})
at $M^B=M^{\omega}=0$.
It enters Eq. (\ref{14}) with an arbitrary weight,
i.e. multiplied by an arbitrary constant $c_M$.
If, following EL approach and DM, we describe the amplitudes
$M_{1+}^{3/2},E_{1+}^{3/2}$ in terms  of the contributions
corresponding to the $N,\Delta,\pi$ and $\omega$ 
exchanges, then $c_MM^{hom}$ should be considered as the
$\Delta$ contribution. In order to obtain the contribution,
corresponing to the $\Delta$ exchange in the $s$-channel,
one should subtract from $c_MM^{hom}$
the contribution of the $\Delta$ exchange in the $u$-channel.
Using final results for the contributions
of $c_MM^{hom}$ into $M_{1+}^{3/2},E_{1+}^{3/2}$,
one can estimate this contribution.
It appeared that the $\Delta$
contribution, corresponding to the $u$-channel,
is negligibly small in comparison with
$c_MM^{hom}$  and $M_{part}^{B,\omega}(W)$.
By this reason, the $\Delta$  contribution in the $s$-channel we identify with
$c_MM^{hom}$.

Let us note, that our final results
correspond to the solutions of the integral equations
(\ref{9})  with $K(W,W')\neq 0$,
i.e. they satisfy the requirement of the crossing invariance.
These solutions were obtained numerically,
using the formulas for the amplitudes
$M_{1+}^{3/2},E_{1+}^{3/2}$
presented in details in Ref. \cite{24}.
At $W_{thr}<W<1.5~GeV$, the phase $\delta_{1+}^{3/2}$
was taken in the analitical form
\begin{equation}
\sin^2\delta ^{3/2}_{1+}=\frac
{(4.27q^3)^2}
{(4.27q^3)^2+(q_r^2-q^2)^2
[1+40q^2
(q^2-q_r^2)+21.4q^2]^2},
\label{100}\end{equation} 
which describe well the experimental data from \cite{31,32}
with $q_r=0.225~GeV$; $q$ is the 3-momentum of the pion
in the $GeV$ units in the $\pi N$ c.m.s.
\section{Results and discussion}
In this Section we present our results on the description
of the  data for the multipole amplitudes
$M_{1+}^{3/2},E_{1+}^{3/2}$
which are extracted with high accuracy
from existing experimental data in the partial- wave
analysis of Ref. \cite{28}.
In the dispersion relation approach, presented in the previous Section,
these data should be described as sums of the particular 
and homogeneous solutions of the integral equations (\ref{9})
for the amplitudes $M_{1+}^{3/2}$ and $E_{1+}^{3/2}$.
The particular solutions have definite magnitudes
fixed by $M^B$ and $M^{\omega}$,
i.e. by the $N,\pi$ and $\omega$ contributions into
$M_{1+}^{3/2},E_{1+}^{3/2}$.
The solutions of the homogeneous parts of the integral
equations (\ref{9}) with $M^B=M^{\omega}$,
have definite shapes, fixed by the integral
equations, and arbitrary weights.
These weights are the only unknown parameters
which should be found from the requirement
of best description of the data on 
$M_{1+}^{3/2},E_{1+}^{3/2}$.
For this aim we have used fitting procedure.

The obtained results together with the data from
Ref.\cite{28} are presented in Figs.1,2.
In order to demonstrate the role of different contributions,
they are presented in these figures separately.

The curves 4 and 6 are the particular solutions of Eg.(\ref{9})
generated by $M^B$ and $M^{\omega}$, respectively.
They represent the nonresonance contributions into
$M_{1+}^{3/2},E_{1+}^{3/2}$,
caused by the  $N,\pi$ and $\omega$ exchanges.
The curves 5 represent the first term in Eq.(\ref{4})
with $M^{NR}=M^B$.
They are given in order to demonstrate the difference between
the nonresonance contributions, generated by the Born term
in the EL aproach of Ref.\cite{13} and our approach.
This difference is caused by the second term
in (\ref{4}); with this term, the nonresonance contributions,
generated by the Born term,
satisfy dispersion relations.

The curves 3 represent the contributions of the homogeneous
solutions, obtained by fitting the weights
of these solutions, when the nonresonance contributions
are generated by the Born term and $\omega$
exchange. These curves represent the $\Delta$
contributions into $M_{1+}^{3/2},E_{1+}^{3/2}$.
As it was mentioned in Sec.2,
our estimations have shown that the $u$-channel
$\Delta$ contributions are negligibly small
in comparison with $s$ -channel ones.
By this reason we identify the contributions of the homogeneous
solutions (curves 3) with the $\Delta$ exchange
in the $s$-channel.

The summary results are presented by the curves 1,
which correspond to the case, when the nonresonance
contributions are caused by the 
$N,\pi$ and $\omega$ exchanges.
It is seen that the agreement with the VPI data
is good for both amplitudes
$M_{1+}^{3/2},E_{1+}^{3/2}$.
In order to demonstrate the role of high energy
contributions into dispersion integrals which are approximated
in our approach by the $\omega$ exchange,
we present also the curves 2. They are obtained by fitting
the homogeneous solutions, when the nonresonance
contributions are generated by the Born terms only.
It is seen that the $\omega$ contribution
is small; however, in the case of $M_{1+}^{3/2}$
its role in obtaining the good agreement with experiment
is important.

In Table 1 we present the helicity amplitudes
$A^p_{3/2}$ and $A^p_{1/2}$ and the ratio $E2/M1$
for the transition $\gamma N \rightarrow P_{33}(1232)$,
which are obtained from the resonance
contributions into $M_{1+}^{3/2},E_{1+}^{3/2}$
(the curves 3 in Figs.1,2) at the resonance position.
First errors are obtained assuming that the data in Figs.1,2,
corresponding to the energy-dependent analysis
of Ref.\cite{28} have $2\%$ errors.
Second errors come from the uncertainties of the model.
They are connected with the cuttof in the dispersion
integrals (\ref{9}); with the uncertainties in the
$\omega$ contribution; with neglecting the couplings
of the multipole amplitudes with each other in (\ref{9});
and with the uncertainties in the extraction
of the resonance amplitudes from the curves 3, discussed
in the previous Section.

Table 1. Helicity amplitudes and the ratio
$E2/M1$ for the $\gamma N \rightarrow P_{33}(1232)$ transition\\
\begin{tabular}{|c|c|c|c|}
\hline
&$A_{1/2}^p(10^{-3}Gev^{-1/2})$&$A_{3/2}^p(10^{-3}Gev^{-1/2})$&$E2/M1(\%)$\\
\hline
Resonance contributions,&$-110\pm 2\pm 6$&$-209\pm 4\pm 12$&$-2.2\pm 0.1\pm 0.3$\\
our results&&&\\
\hline
Total amplitudes,&$-135\pm 5$&$-250\pm 8$&$-1.5\pm 0.5$\\
Ref.\cite{28}&&&\\
\hline
Nonrelativistic&-101&-175&0\\
quark model&&&\\
\hline
Relativistic&-111&-207&-2.1\\
quark model \cite{36,37}&&&\\
\hline
\end{tabular} 
\newline
 
In Table 1 we present also the results obtained from the total
amplitudes  $M_{1+}^{3/2},E_{1+}^{3/2}$
at the resonance position in Ref. \cite{28}.
The amplitudes, extracted in such way, are larger
than quark model predictions.
As is seen from our results, this disagreement
is removed due to taking into account the nonresonance
background contributions generated by the
$N,\pi$ and $\omega$ exchanges.
\newline
\begin{center}
{\large {\bf {Acknowledgments}}}
\end{center}

I am grateful to I.I.Strakovsky for communications and providing
the results of the VPI partial-wave analysis  in the numerical form.
I also acknowledge communications with B.L.Ioffe and O.Hanstein.  
 
\newpage
{\Large \bf {Figure Captions}}
\vspace{1cm}
\newline
{\large \bf{Fig. 1}}  Multipole amplitude $M_{1+}^{3/2}$.
Our results for the imaginary parts of the amplitude (curve 1)
in comparison with VPI data \cite{28}:
the solid circles represent the results of 
the energy-dependent analysis, the open circles
correspond to the energy-independent analysis.
Curve 3 correspond to the resonance contribution.
Other contributions are discussed in the text. 
\newline
{\large \bf{Fig. 2}} Multipole amplitude $E_{1+}^{3/2}$.
 The legend is as for Fig.1.

\vspace{1cm}

\begin{thebibliography}{999}

\bibitem{1} W.Pfeil, D.Schwela, Nucl. Phys. $\bf{B45}$, 379 (1972).
\bibitem{2} F.A.Berends, A.Donnachie,Nucl. Phys. $\bf{B84}$, 342 (1974). 
\bibitem{3} M.G.Olsson, Nucl.Phys. $\bf{B78}$, 55 (1974).
\bibitem{4} M.G.Olsson, Phys.Rev. D $\bf{13}$, 2502 (1976).
\bibitem{5} M.G.Olsson, Nuovo Cim. $\bf{A40}$, 284 (1977).
\bibitem{6} K.M.Watson, Phys. Rev. $\bf{95}$, 228 (1954).
\bibitem{7} A.M.Bernstein, S.Nozawa, M.A.Moienster, Phys.Rev. C $\bf{47}$, 1274 (1993).
\bibitem{8} A.S.Omelaenko, P.V.Sorokin, Sov. J. Nucl. Phys. $\bf{38}$, 398 (1983).
\bibitem{9} R.M.Davidson, N.C.Mukhopadhyay, Phys.Rev. D $\bf{42}$, 20 (1990).  
\bibitem{10} M.G.Olsson, E.T.Osypowski, Nucl.Phys. $\bf{B87}$, 399 (1975).
\bibitem{11} M.G.Olsson, E.T.Osypowski, Phys.Rev. D $\bf{17}$, 174 (1978).
\bibitem{12} R.M.Davidson, N.C.Mukhopadhyay, R.Wittman, Phys.Rev.Lett. $\bf{56}$, 804 (1986).
\bibitem{13} R.M.Davidson, N.C.Mukhopadhyay, R.Wittman, Phys.Rev. D $\bf{43}$, 71 (1991).
\bibitem{14} J.M.Laget, Nucl.Phys. $\bf{A481}$, 765 (1988).
\bibitem{15} H.Tanabe, K.Ohta, Phys.Rev. C $\bf{31}$, 1876 (1985).
\bibitem{16} S.N.Yang, J.Phys. G $\bf{11}$, L205 (1985).
\bibitem{17} S.Nozawa, B.Blankleider, T.-S.H.Lee, Nucl.Phys. $\bf{A513}$, 459 (1990).
\bibitem{18} S.Nozawa, B.Blankleider, T.-S.H.Lee, Nucl.Phys. $\bf{A513}$, 511 (1990).
\bibitem{19} T.-S.H.Lee, B.C.Pearce, Nucl.Phys. $\bf{A530}$, 532 (1991).
\bibitem{20} T.Sato,T.-S.H.Lee, Phys.Rev. C $\bf{54}$, 2660 (1996).
\bibitem{21} Y.Surya, F.Gross, Phys.Rev. C $\bf{53}$, 2422 (1996).	
\bibitem{22} O.Hanstein, D.Drechsel, L.Tiator, Phys. Lett. $\bf{385B}$, 45 (1996).
\bibitem{23} O.Hanstein, D.Drechsel, L.Tiator, Nucl.Phys. $\bf{A632}$, 561 (1998).
\bibitem{24} I.G.Aznauryan, Phys.Rev. D $\bf{57}$, 2727 (1998).
\bibitem{25} I.G.Aznauryan, S.G.Stepanyan, Phys.Rev. D $\bf{59}$, 054009 (1999) .
\bibitem{26} D.Schwela, H.Rolnik, R.Weizel, W.Korth, Z. Phys. $\bf{202}$, 452 (1967).
\bibitem{27} D.Schwela, R.Weizel, Z. Phys. $\bf{221}$, 71 (1969).
\bibitem{28} R.A.Arndt, I.I.Strakovsky, R.L.Workman, Phys.Rev. C $\bf{56}$, 577 (1997).
\bibitem{29} P.Noelle, Prog.Theor.Phys. $\bf{60}$, 778 (1974).
\bibitem{30} M.L.Goldberger, K.M.Watson, Collision Theory, 
(John Wiley $\&$ Sons, Inc., New York-London-Sydney, 1964).
\bibitem{31} R.A.Arndt, I.I.Strakovsky, R.L.Workman, M.M.Pavan, 
 Phys.Rev. C $\bf{52}$, 2120 (1995).
\bibitem{32} R.A.Arndt, I.I.Strakovsky, R.L.Workman, M.M.Pavan,
e-print, nucl-th/9807087.
\bibitem{33} R.Worden, Nucl.Phys. $\bf{B37}$, 253 (1972).
\bibitem{34} Review of Particle Physics, Phys.Rev. D $\bf{54}$ (1996).
\bibitem{35} O.Dumbrajs, R.Koch, P.Pilkuhn et al, Nucl.Phys. $\bf{B216}$, 277 (1983).
\bibitem{36} I.G.Aznauryan, Phys. Lett. $\bf{316B}$, 391 (1993).
\bibitem{37} I.G.Aznauryan, Z. Phys. A $\bf{346}$, 297 (1993).
\end{thebibliography}
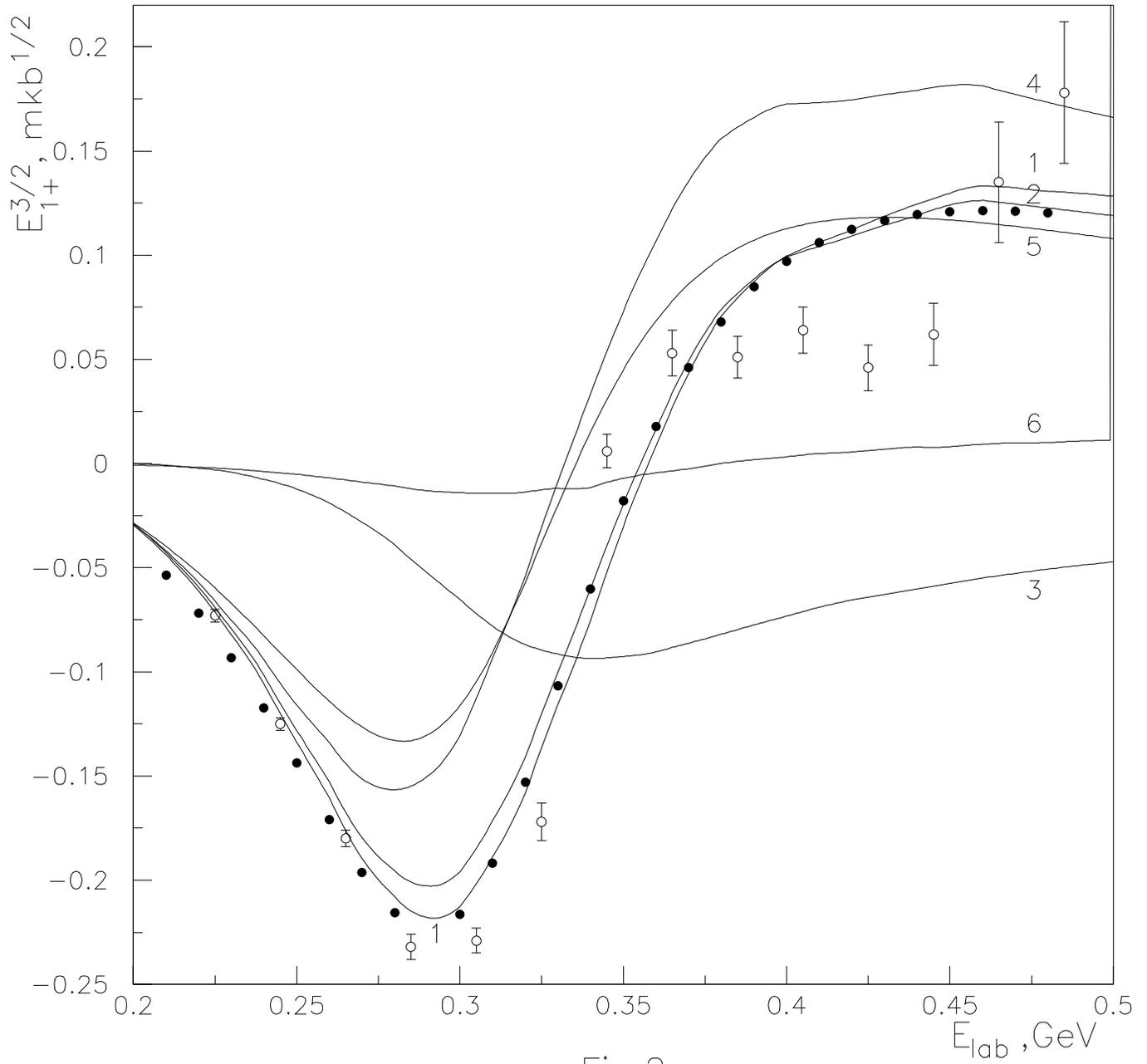
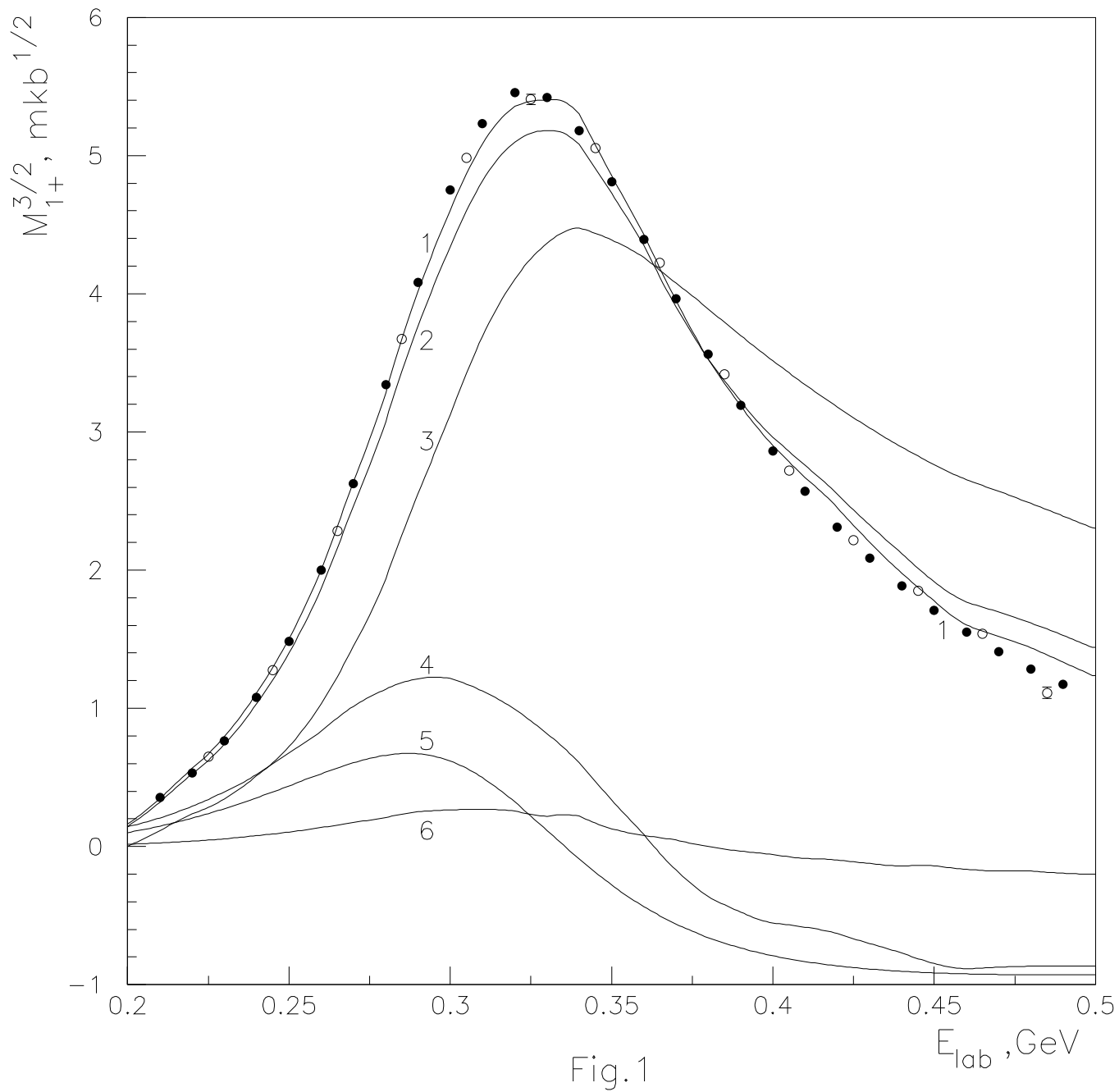
\end{document}